\documentclass[reprint,aps,prl,twocolumn,superscriptaddress,nofootinbib]{revtex4-2}

\usepackage{amsmath,amssymb,bm,graphicx}
\usepackage[dvipsnames]{xcolor}
\usepackage{times}
\usepackage[colorlinks=true,linkcolor=MidnightBlue,citecolor=MidnightBlue,urlcolor=MidnightBlue]{hyperref}

\begin{document}
\title{Interaction-Induced Quasicrystalline Order: Emergence of Quasi-Solid and Quasi-Supersolid Phases}

\author{Chao Zhang}
\email{chaozhang@ahnu.edu.cn}
\affiliation{Department of Physics, Anhui Normal University, Wuhu, Anhui 241000, China}

\begin{abstract}
Deterministic quasiperiodicity in quantum systems has long been associated with localization, criticality, or glassy behavior, and has therefore been believed to suppress long-range order rather than stabilize it. Here we demonstrate the opposite: quasiperiodicity in interactions—without any quasiperiodic potential, disorder, or geometric modulation—can generate coherent, ordered quantum phases. We study hard-core bosons in one dimension with quasiperiodic long-range interactions, $V_{ij}=V_0 \cos(\pi \alpha i)\cos(\pi \alpha j)$, where $\alpha=(\sqrt{5}-1)/2$ is the inverse golden ratio. Using large-scale path-integral quantum Monte Carlo simulations, we uncover thermodynamically stable incompressible plateaus at irrational densities tied to Fibonacci ratios. These plateaus exhibit sharp incommensurate Bragg peaks, signaling an emergent quasi-solid with long-range quasicrystalline density order. More strikingly, at nearby fillings and interaction strengths, we identify a quasi-supersolid phase that supports both Fibonacci density ordering and finite superfluid density—demonstrating that interaction-induced quasiperiodicity can stabilize supersolid coherence. Our results establish a new mechanism for realizing ordered quasicrystalline quantum matter, and provide realistic guidance for implementation in Rydberg atom arrays, multimode cavity-QED systems, and trapped-ion quantum simulators.
\end{abstract}

\maketitle

\noindent \textbf{Introduction--} Quasiperiodic quantum systems sit at the intriguing interface between order and disorder: deterministic yet aperiodic, neither random nor periodic, they challenge the conventional paradigms of quantum matter. Over the past four decades, quasiperiodicity has become synonymous with localization~\cite{aubry1980analyticity,SanchezPalencia2010_NatPhys}, criticality~\cite{hiramoto1989scaling,jagannathan2021review,sun2024fibonacci}, and Bose-glass phases~\cite{Roux2008_PRA, Deng2008_PRA, Larcher2009_PRA, Zhang2015, g7vd-hgw4, PhysRevB.110.125124}, where long-range coherence is suppressed rather than stabilized. From Aubry--Andr\'e potentials and bichromatic lattices to Fibonacci sequences and quasicrystalline geometries, studies have consistently concluded that quasiperiodicity destroys order---not creates it~\cite{aubry1980analyticity,SanchezPalencia2010_NatPhys,Roux2008_PRA,Deng2008_PRA,Larcher2009_PRA,Zhang2015,g7vd-hgw4,PhysRevB.110.125124,PhysRevB.96.035153,hiramoto1989scaling,jagannathan2021review,mace2019many,varma2019diffusive,strkalj2021interpolating,schreiber2015science,kohlert2019prl,zhang2018mblqp}.

\emph{But is this always true? Can quasiperiodicity, under the right conditions, generate---rather than suppress---quantum order?}  
This core question has remained essentially unexplored because quasiperiodicity has almost exclusively been implemented through \emph{external on-site potentials} or \emph{lattice geometry}, rather than through the \emph{interactions themselves}. Potential-driven quasiperiodicity inherently competes with coherence, favoring localization, multifractality, or glassy behavior, even in strongly interacting regimes. What has been largely overlooked is a fundamentally different mechanism: \emph{deterministic quasiperiodicity embedded directly in the interaction kernel}, instead of the on-site potential.

Recent theoretical work has proposed that quasiperiodic long-range interactions can be engineered, for example, using multimode cavity interference, yielding real-space interaction profiles that approximate Fibonacci-like modulations even in the absence of external quasiperiodic potentials~\cite{PhysRevA.106.023307}. At the same time, rapid advances in programmable quantum simulators---including Rydberg-atom arrays with tunable blockade radii~\cite{scholl2021,browaeys2020}, cavity-mediated atomic ensembles with controllable mode structures~\cite{PhysRevX.8.011002,landig2016}, and trapped-ion platforms with engineered power-law interactions~\cite{RevModPhys.93.025001}---have brought the physical realization of interaction-engineered models within reach. These developments naturally raise a profound and previously unresolved question: \emph{Can deterministic quasiperiodicity in the \emph{interaction sector alone} stabilize ordered quantum phases---including solids or supersolids---that have long been believed incompatible with quasiperiodic systems?}

In this work, we answer this question in the affirmative. We investigate a one-dimensional hard-core Bose--Hubbard model with \emph{purely interaction-induced quasiperiodicity}, defined by $V_{ij} = V_0 \cos(\pi \alpha i)\cos(\pi \alpha j),$ where $\alpha = (\sqrt{5}-1)/2$ is the inverse golden ratio. Crucially, our model contains neither on-site potential disorder nor quasiperiodic on-site modulation, thereby isolating the role of quasiperiodic long-range interactions alone.

Using large-scale path-integral quantum Monte Carlo simulations with the worm algorithm~\cite{Prokofev1998_JETP,Prokofev1998_PLA}, we uncover \emph{two thermodynamically stable quantum phases that, to the best of our knowledge, have not been observed in any previous quasiperiodic system}:
(i) a \emph{quasi-solid} phase, marked by incompressible density plateaus at Fibonacci fillings and sharp incommensurate peaks in the structure factor, and
(ii) a \emph{quasi-supersolid} phase, where such quasicrystalline density modulations coexist with finite superfluid density.
These results demonstrate that quasiperiodicity in the interaction kernel---rather than suppressing coherence---can drive the emergence of ordered and even supersolid quantum matter, overturning the conventional belief that quasiperiodicity inevitably leads to glassiness or localization.

Moreover, because the required quasiperiodic interaction patterns can be engineered using Rydberg-atom arrays, multimode cavity-QED systems, and trapped-ion simulators, our findings provide a realistic roadmap for realizing and probing quasicrystalline solids and supersolids in next-generation quantum simulators. In this sense, interaction-driven quasiperiodicity defines a new paradigm for engineering exotic ordered states beyond conventional supersolids.


\noindent \textbf{Model--}
We consider a one-dimensional system of hard-core bosons described by an extended Bose–Hubbard Hamiltonian with long-range, quasiperiodically modulated interactions,
\begin{equation}
H = -t \sum_{\langle i,j \rangle} \left( a_i^\dagger a_j + \text{h.c.} \right)
- \mu \sum_i n_i
+ \sum_{i \ne j} V_{ij} n_i n_j ,
\label{eq:H}
\end{equation}
where $a_i^\dagger (a_i)$ creates (annihilates) a boson at site $i$, and $n_i = a_i^\dagger a_i$ is the density operator. We set $t=1$ as the energy unit. The long-range interaction takes the form $V_{ij} = V \cos(\pi \alpha i)\cos(\pi \alpha j)$, with strength $V$ and irrational modulation parameter $\alpha = (\sqrt{5}-1)/2$, the inverse golden ratio.

To faithfully capture the self-similar structure of the quasiperiodic modulation, we choose system sizes $L = F_k$ following the Fibonacci sequence defined by $F_0 = 0$, $F_1 = 1$, and $F_k = F_{k-1} + F_{k-2}$. In particular, we use $L = 233 = F_{13}$, which accurately represents the incommensurate modulation through its rational approximant $\alpha \simeq F_{12}/F_{13} = 144/233$. This construction naturally embeds the golden-ratio spatial pattern into the finite lattice and determines the characteristic incommensurate wavevectors observed in statics structure factor $S(q)$. Larger system sizes up to $L = 987$ are also simulated to assess finite-size effects.

\noindent\textbf{Methods and Observables--} We perform large-scale path-integral quantum Monte Carlo simulations using the worm algorithm~\cite{Prokofev1998_JETP, Prokofev1998_PLA}, simulating systems with up to $L = 987$ sites and inverse temperature $\beta = L$ to ensure ground-state convergence. 

To identify different phases, we compute the following observables. The average filling is defined as $\langle n \rangle = \frac{1}{L}\sum_x \langle n_x \rangle$, and incompressible plateaus in $\langle n \rangle(\mu/V)$ indicate gapped states. The compressibility is $\kappa = \frac{\beta}{L}(\langle N^2 \rangle - \langle N \rangle ^2)$, with $N$ total number of particles of the system, estimated via density fluctuations, with vanishing $\kappa$ marking insulating behavior. The superfluid density is $\rho_s = \langle W^2 \rangle/(2 \beta t)$~\cite{PollockCeperley1987_PRB}, where $W$ is the winding number of worldlines. Here, $\rho_s > 0$ implies global phase coherence.

To characterize spatial ordering we compute the imaginary-time–averaged connected density–density correlator 
$C(x)=\frac{1}{L} \sum_{i} \big{(} \langle n_i n_{i+x}\rangle-\langle n_i\rangle\langle n_{i+x}\rangle \big{)} - \langle n \rangle$.
Its Fourier transform, the static structure factor, 
$S(q)=\frac{1}{L}\sum_{x} e^{iqx} C(x)$, 
provides momentum-resolved information on density modulations. 
In periodic systems $S(q)$ exhibits Bragg peaks at commensurate momenta, while in quasiperiodic systems sharp incommensurate peaks signal quasi–long-range order. 
In our model these peaks appear at golden–ratio–related wavevectors $q^*$, offering a clear signature of quasi-solid or quasi-supersolid phases stabilized by interaction-driven quasiperiodicity. We further cross-check the ordering wavevectors by performing an independent Fourier analysis of the (imaginary-time--averaged) real-space density profile,
$\langle n_x \rangle \equiv \frac{1}{\beta}\int_0^\beta d\tau\, \langle n_x(\tau)\rangle$.
Specifically, we compute
$\tilde n(q)=\frac{1}{L}\sum_x e^{iq x}\big(\langle n_x \rangle -\langle n\rangle\big)$.
The dominant peaks in $\tilde n(q)$ occur at the same incommensurate momenta as those in $S(q)$, confirming that the peaks reflect genuine density modulations rather than artifacts of a particular estimator.

With these definitions in hand, we now turn to the central question of how quasiperiodically modulated long-range interactions reshape the many-body phase diagram. In the following section, we present the numerical results obtained from large-scale QMC simulations and discuss their physical implications.

\begin{figure}[t]
    \centering
    \includegraphics[width=\linewidth]{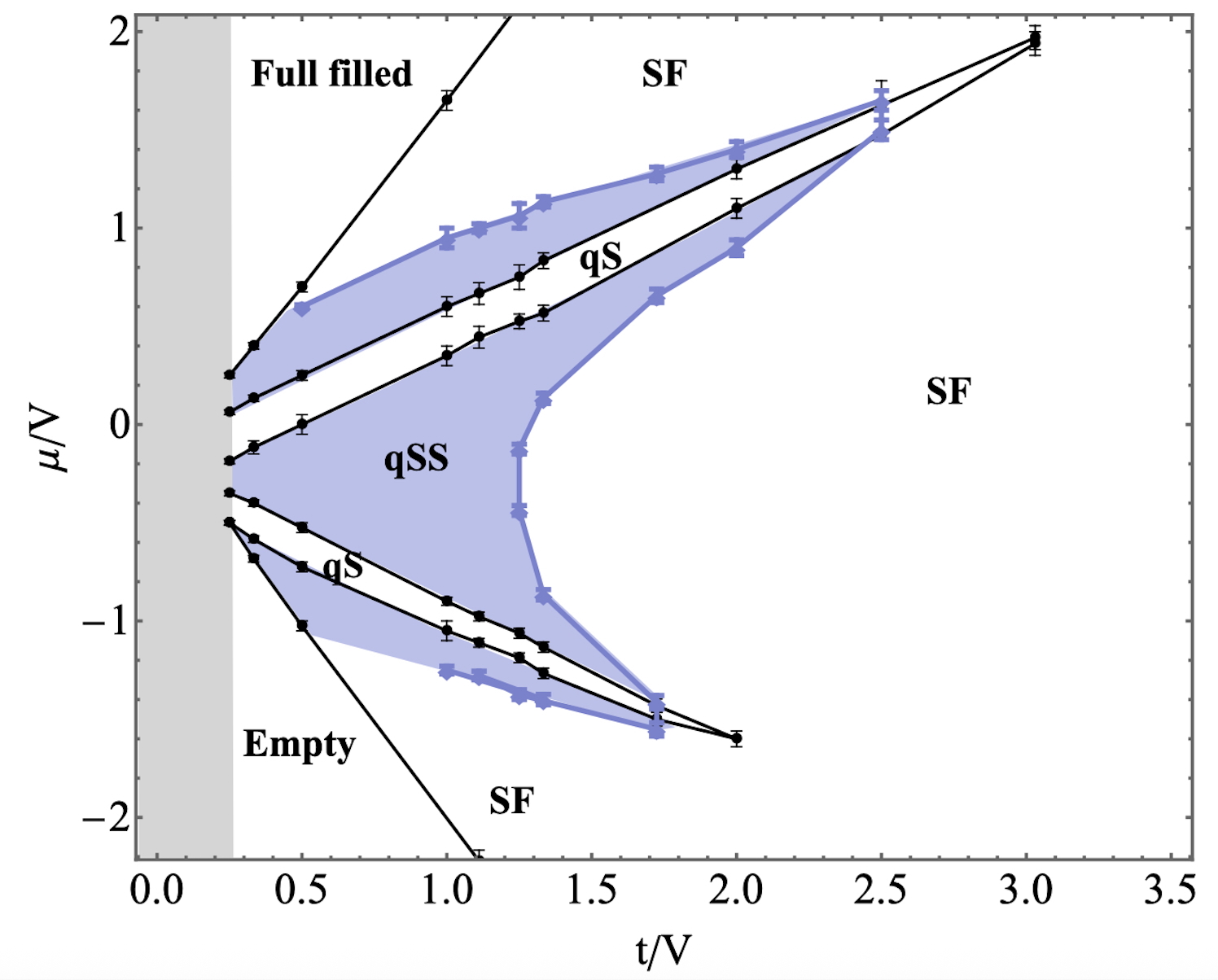}
\caption{
Ground-state phase diagram of the system in the $\mu/V$–$t/V$ plane for $L=233$. 
Two incompressible quasi-solid (qS) lobes emerge near fillings $\langle n \rangle =0.618 \approx \alpha $ and $\langle n \rangle =0.382 \approx 1 - \alpha $, where $\alpha = (1-\sqrt{5})/2$ is the inverse golden ratio. 
These phases exhibit vanishing compressibility and superfluid density, along with sharp incommensurate peaks in $S(q)$, indicating Fibonacci-induced spatial order. 
The light blue region marks the quasi-supersolid (qSS) regime, where incommensurate density modulations coexist with finite superfluid density.
}
    \label{fig:phase_diagram}
\end{figure}

\noindent
\textbf{Phase diagram--}
Figure~\ref{fig:phase_diagram} summarizes the ground-state phase diagram of a one-dimensional hard-core Bose system with Fibonacci-modulated quasiperiodic long-range interactions. 
For system size $L=233$, we map out the phases in the $\mu/V$–$t/V$ plane and identify two insulating lobes near fillings $\langle n \rangle \approx \alpha$ and $1 - \alpha$, where $\alpha$ is the inverse golden ratio. 
These lobes exhibit zero compressibility ($\kappa=0$), vanishing superfluid density ($\rho_s=0$), and sharp incommensurate Bragg peaks in $S(q)$—hallmarks of \emph{quasi-solid} phases stabilized solely by deterministic incommensurate interactions.

Unlike Mott or charge-density-wave insulators, which rely on on-site interactions or spatially periodic potentials, the ordering here emerges without any external modulation. 
Instead, the quasiperiodicity is encoded directly in the interaction profile, driving incommensurate density locking at irrational fillings. 
This mechanism circumvents conventional commensuration and leads to robust, aperiodic solids of purely interaction-driven origin.

Upon increasing $\mu/V$ at fixed $t/V$, the system undergoes melting transitions into intermediate regions with coexisting density modulation and finite $\rho_s$, indicating the formation of \emph{quasi-supersolid} (qSS) phases. 
These qSS regions persist over a wide parameter range and arise from the interplay of quasiperiodic long-range repulsion and quantum fluctuation, without the need for soft-core interaction or potential-based mechanisms.

As $t/V$ increases, kinetic energy suppresses the quasiperiodic ordering, shrinking both the quasi-solid lobes and quasi-supersolid regions. 
At large hopping, the system enters into a compressible superfluid (SF) phase with uniform density and finite superfluid density. 
The resulting phase diagram thus reveals a rare sequence: quasi-solid $\to$ quasi-supersolid $\to$ SF, driven entirely by the competition between deterministic quasiperiodic long-range interactions and quantum fluctuations.

\begin{figure}[t]
    \centering
    \includegraphics[width=\linewidth]{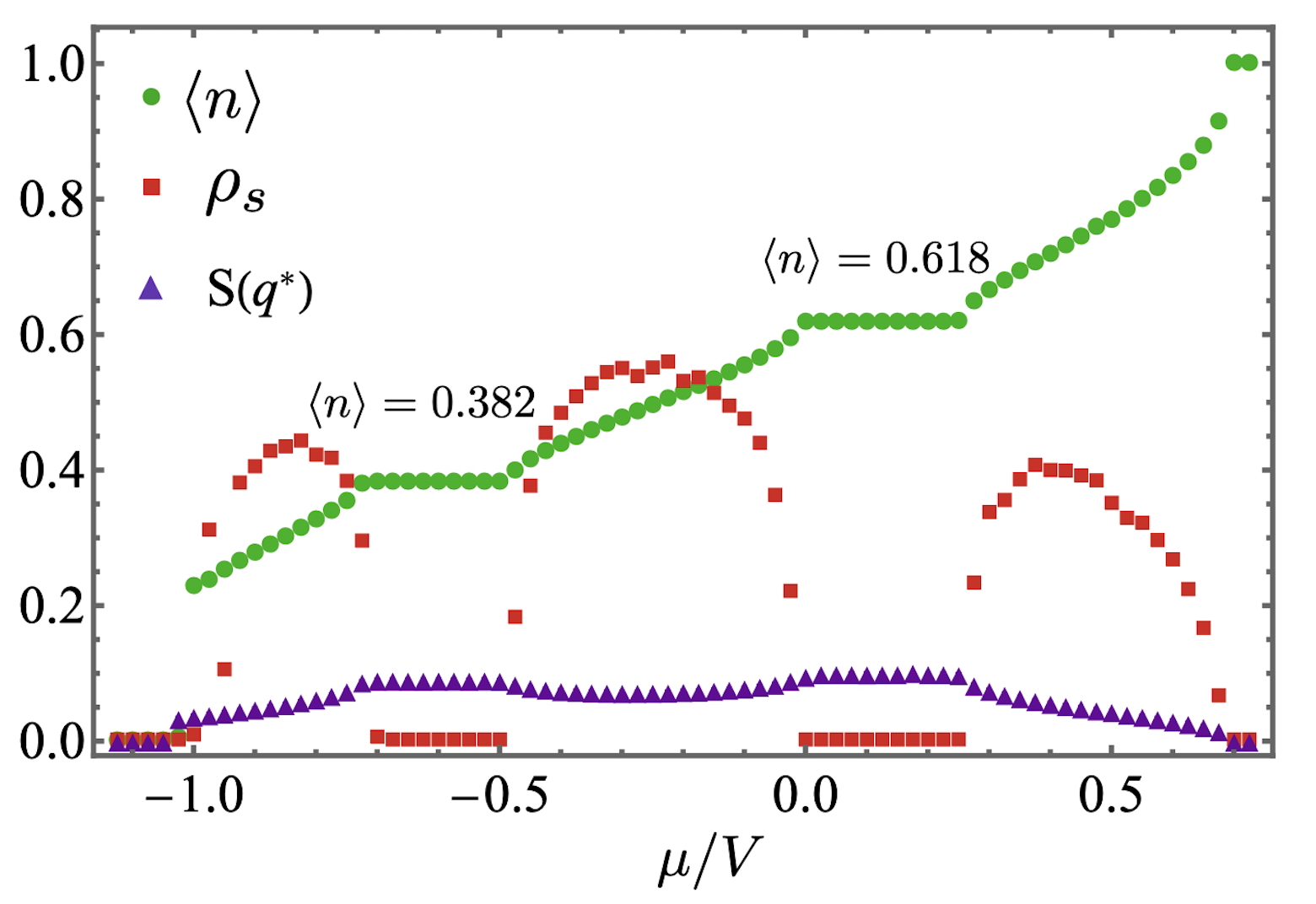}
\caption{
Average filling $\langle n \rangle$ (green circles), superfluid density $\rho_s$ (red squares), 
and structure factor at the dominant incommensurate peak $S(q^*)$ (purple upward triangles) 
as functions of chemical potential $\mu/V$ for $L=233$ at $t/V=0.5$. 
Two prominent incompressible plateaus appear at irrational fillings 
$\langle n \rangle = 0.618$ and $\langle n \rangle =0.382$, corresponding to the inverse golden-ratio 
$\alpha$ and $1 - \alpha$ with $\alpha = (1-\sqrt{5})/2$. 
These densities are naturally commensurate with the underlying Fibonacci-modulated 
interaction pattern and host quasi-solid order, characterized by vanishing 
$\rho_s$ and enhanced $S(q^*)$. 
Between these plateaus, finite $\rho_s$ coexists with nonzero $S(q^*)$, 
signaling the emergence of a quasi-supersolid phase stabilized purely by 
quasiperiodic long-range interactions. Error bars are within symbols if not seen in the figure.
}
    \label{fig:mu_scan}
\end{figure}

To characterize the quasi-solid phase, Fig.~\ref{fig:mu_scan} presents the average filling $\langle n \rangle$, superfluid density $\rho_s$, and the incommensurate peak height of the structure factor $S(q^*)$ at $q^* = 2.398$, plotted as functions of the chemical potential $\mu/V$ for fixed $t/V = 0.5$ and system size $L = 233$. Two pronounced incompressible plateaus emerge at irrational fillings $\langle n \rangle=0.382$ and $\langle n \rangle =0.618$, directly associated with Fibonacci ratios. At these fillings, $\rho_s$ vanishes while $S(q^*)$ is strongly enhanced, signaling the stabilization of a \emph{quasi-solid} phase characterized by incommensurate density order without periodicity.  In addition, an incompressible lobe appears at filling $\langle n \rangle=1.0$, resulting from the hard-core constraint, which enforces a fully occupied configuration with no available fluctuations.

Crucially, the regions between the incompressible lobes do not simply represent a superfluid phase. Instead, we observe a regime where finite superfluid density $\rho_s$ coexists with pronounced incommensurate density modulations, as signaled by a non-vanishing structure factor $S(q^*) > 0.03$. This threshold is determined by analyzing the background value of $S(q^*)$ when the quasi-solid order disappears (see Supplemental Material). The coexistence of off-diagonal long-range coherence and golden-ratio-related density modulation identifies this intermediate phase as a \textit{quasi-supersolid}. It emerges purely from quasiperiodic long-range interactions, without the need for external lattice commensuration, on-site disorder, or on-site/nearest-neighbor interactions. This mechanism offers a novel route to supersolidity driven entirely by deterministic interaction patterns. The extended stability of these quasi-supersolid regimes between robust Fibonacci-related solids highlights the capability of interaction-driven quasiperiodicity to stabilize exotic many-body phases in low-dimensional quantum systems.

\begin{figure}[t]
\centering
\includegraphics[width=1.0\linewidth]{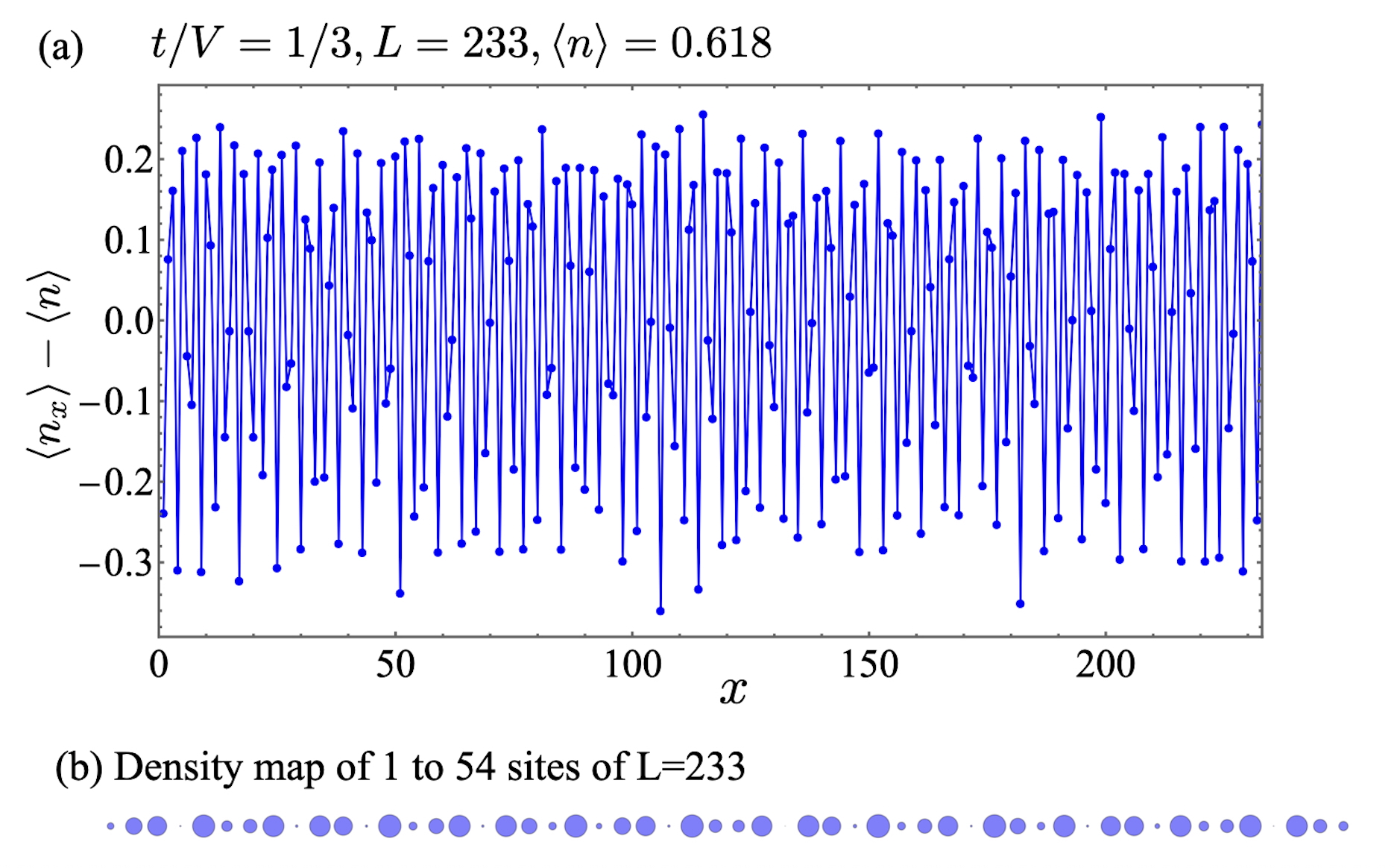}
\caption{(a) Site-resolved average density profile \(\langle n_x \rangle -\langle n \rangle \) along a one-dimensional chain of length \(L=233\), with total particle number \(N=144\), corresponding to filling \( \langle n \rangle=0.618 \approx \alpha\), where \(\alpha = (1-\sqrt{5})/2\) is the inverse golden ratio at $t/V=1/3$. The quasiperiodic long-range interaction follows a Fibonacci modulation pattern, leading to a quasi-solid phase characterized by incommensurate density modulations, vanishing superfluid density, and zero compressibility.  
(b) Zoom-in of sites 1–54 of the same system. The circle radii are proportional to the local density at each site, emphasizing the incommensurate spatial modulation.}
\label{fig:density_profile}
\end{figure}

In Fig.~\ref{fig:density_profile}(a), we show the spatial distribution of the averaged density profile \(\langle n_x \rangle - \langle n \rangle \) for \(L=233\) and total number of particles in the system \(N=144\) for hard-core bosons. The system is deep in the correlated regime ($t/V=1/3$), where both compressibility and superfluid density vanish. Instead of a uniform profile or a commensurate charge-density wave (CDW), the density exhibits robust incommensurate oscillations reflecting the Fibonacci modulation of the interaction. The modulation is present throughout the lattice, indicating robust incommensurate order at system size $L=233$.  

Importantly, the site occupations remain non-integer, taking values between approximately 0.4 and 0.8, rather than pinning to 0 or 1 as in conventional CDW states, as shown in Fig.~\ref{fig:density_profile}(b). This feature signals the formation of a \emph{quasi-solid} phase: globally incompressible, yet locally fluid-like, with continuous spatial variations in density. 

Unlike crystalline order in periodic systems, the quasi-solid state inherits deterministic but aperiodic modulations from the long-range Fibonacci couplings. This mechanism stabilizes density order without external potential modulation or randomness, distinguishing it sharply from both CDWs and disorder-driven insulating states. As discussed below, the quasi-solid behavior is further corroborated by the correlation function \(C(x)\) and structure factor \(S(q)\), which reveal sharp Bragg-like peaks at incommensurate momenta.

\begin{figure}[t]
    \centering
    \includegraphics[width=\linewidth]{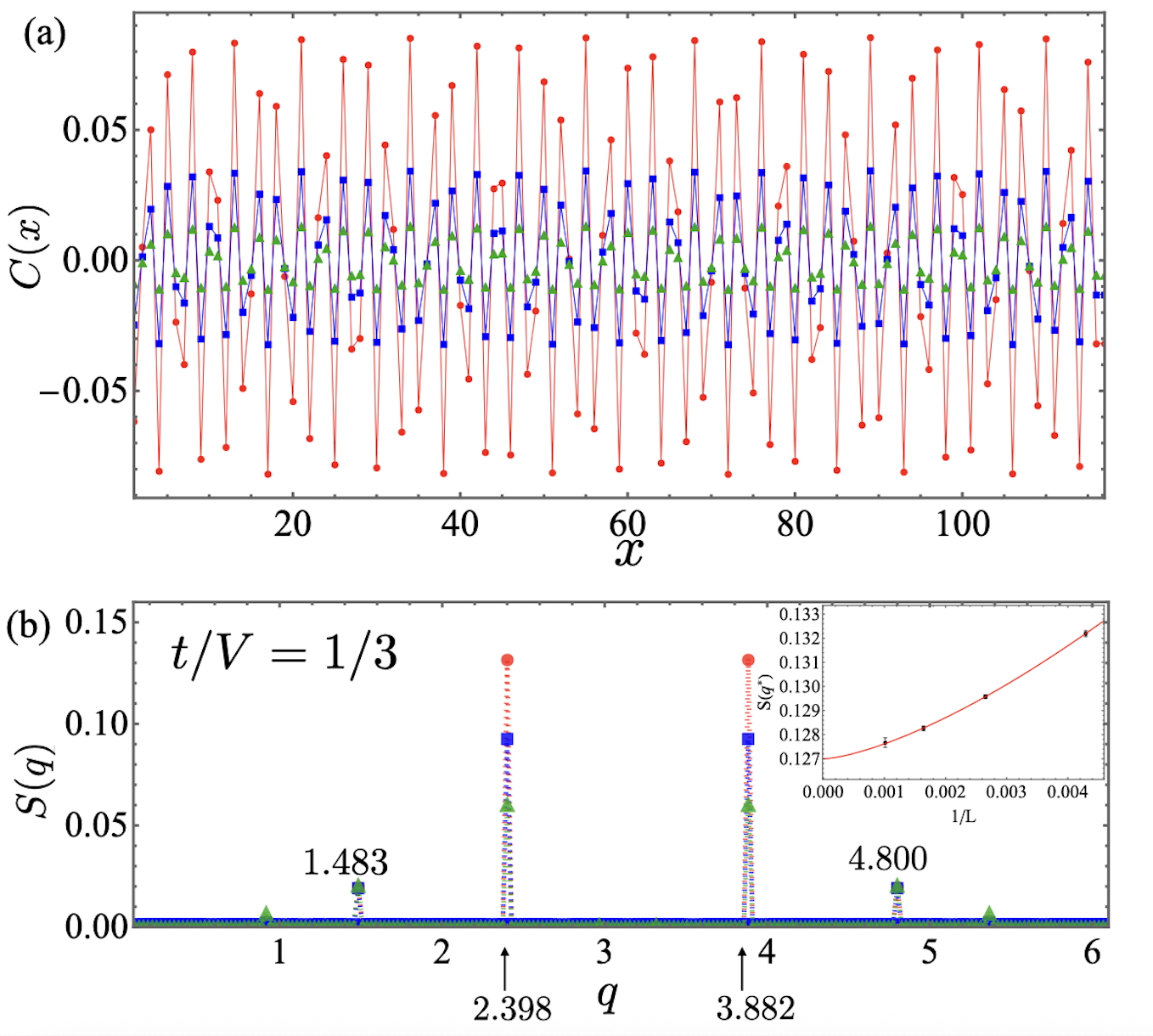}
    \caption{
    (a) Density–density correlation function $C(x)$ at $t/V=1/3$ for three representative fillings on a system of size $L=233$: $\langle n \rangle=0.618$ (red circles), $\langle n \rangle=0.694$ (blue squares), and $\langle n \rangle =0.780$ (green upward triangles). Red data correspond to the inverse golden ratio $\langle n \rangle =0.618 \approx \alpha $, at which quasiperiodic modulation is most pronounced. Fitting curves are superimposed on the data.  
    (b) Corresponding structure factor $S(q)$ for each filling, displaying prominent peaks at incommensurate wavevectors $q_1 = 2.398$ and $q_2 = 3.882$ that reflect the dominant Fourier components of the Fibonacci pattern. Inset: finite-size scaling of the dominant peak height $S(q^*)$ at $q^*=2.398$ for $\langle n \rangle= 0.618$ with the fitting function as $S(q^*)(L)= 0.126865 + 12.3117\,\left(\frac{1}{L}\right)^{1.42065}$, which confirms the persistence of quasi-long-range order in the thermodynamic limit.   }
    \label{fig:correlation_fft}
\end{figure}

To further characterize the nature of the emergent quasi-solid and quasi-supersolid phases, we analyze the time averaged connected density–density correlation function $C(x) = \frac{1}{L} \sum_i (\langle n_i n_{i+x} \rangle - \langle n_i \rangle \langle n_{i+x} \rangle) -\langle n \rangle$ and its Fourier transform $S(q)$ at $t/V = 1/3$ for system size $L = 233$. Figure~\ref{fig:correlation_fft}(a) shows $C(x)$ at three representative fillings: $\langle n \rangle=0.618$, $0.694$, and $0.780$. At $\langle n \rangle = 0.618 \approx \alpha$, $C(x)$ displays coherent quasiperiodic oscillations over long distances, in stark contrast to the more damped behavior at nearby fillings. We fit $C(x)$ using the function $C(x) = A \cos(q x) e^{-x/\xi} + C_0$ for each case. For $\langle n \rangle=0.618$, the best-fit parameters are $A = 0.0837$, $q = 2.4$, $\xi \approx 185001$, and $C_0 \approx -2.34 \times 10^{-7}$; for $\langle n \rangle=0.694$, the fitting yields $A = 0.0334$, $q = 2.40001$, $\xi \approx 31615.1$, and $C_0 \approx -5.34 \times 10^{-7}$; while for $\langle n \rangle=0.780$, we obtain $A = 0.0122$, $q = 2.40001$, $\xi \approx 4962.7$, and $C_0 \approx -1.21 \times 10^{-6}$. The gradual reduction in both the amplitude $A$ and correlation length $\xi$ as the filling deviates from the Fibonacci value does not imply the disappearance of ordered phases. Rather, it reflects a smooth transition from a strongly correlated \textit{quasi-solid} state at $\langle n \rangle=0.618$—where density modulations dominate and $\rho_s=0$—to a \textit{quasi-supersolid} regime at nearby incommensurate fillings. In this quasi-supersolid regime, the modulated density pattern persists while phase coherence emerges, as evidenced by finite values of superfluid density $\rho_s = 0.297 \pm 0.0008$, compressibility $\kappa = 0.323 \pm 0.007$ at $\langle n \rangle = 0.694$, and $\rho_s = 0.247 \pm 0.0003$, $\kappa = 0.384 \pm 0.009$ at $\langle n \rangle = 0.780$. This evolution highlights the robustness of quasiperiodic interaction–induced ordering: while the strength of density modulation weakens away from the Fibonacci resonance, the system continuously retains quasi-supersolid phase at $t/V=1/3$.

The Fourier spectrum shown in Fig.~\ref{fig:correlation_fft}(b) exhibits sharp Bragg‑like peaks at \( q_1 = 2.398 \) and \( q_2 = 3.882 \), in excellent agreement with the expected quasiperiodic wavevectors \( q = 89 \times 2\pi / 233 \) and its symmetry partner \( 2\pi - q \). The persistence of these peaks under finite‑size scaling [see Figure~\ref{fig:correlation_fft} (b) inset] demonstrates that the density modulations form a \textit{quasi‑solid} state with quasi‑long‑range order (for more details in supplemental material).  

For a system of length \( L = F_k \), where \( F_k \) denotes the \( k \)‑th Fibonacci number, the leading Fourier components of the quasiperiodic modulation occur at $q^* = 2 \pi \frac{F_{k-2}}{F_k}$ and its mirror \( 2\pi - q^* \). These characteristic momenta arise from the Fourier transform of the Fibonacci characteristic sequence and are a hallmark of its singular‑continuous diffraction spectrum~\cite{Baake2012, Wolny2000, Xu2003}. In our case with \( L = 233 = F_{13} \), the dominant peak appears at \( q^* = 2.398 \), corresponding to \( 2\pi \times 89 / 233 \), confirming that the observed density pattern is intrinsically locked to the inverse golden ratio \( \alpha = (1 - \sqrt{5})/2 \) through its rational approximants \( F_{k-2} / F_{k} \). The self‑similar nature of the Fibonacci sequence therefore governs the emergence of quasi‑long‑range spatial order. In addition to these principal peaks, two weaker features appear at \( q_3 = 1.483 \) and \( q_4 = 4.800 \). These originate from higher‑order Fourier components of the singular‑continuous spectrum — specifically from secondary harmonics associated with other Fibonacci ratios such as \( F_{k-3}/F_k \), and reflect the hierarchical, self‑similar structure of the underlying quasiperiodic modulation.

Upon doping away from $\langle n \rangle =\alpha$, the Bragg peaks gradually decreases while superfluid density remains finite, signaling a transition to a quasi-supersolid regime. This demonstrates that quasiperiodic long-range interactions can stabilize both incompressible quasi-solids and nearby quasi-supersolid phases with simultaneous density order and phase coherence.

\noindent\textbf{Discussion and Outlook--} Our results demonstrate that deterministic quasiperiodic long-range interactions—without any external potentials, disorder, or soft-core effects—are sufficient to stabilize two exotic many-body phases: the quasi-solid and the quasi-supersolid. These ordered phases emerge solely from the incommensurate structure encoded in the interaction profile and are characterized by the coexistence of spatial density modulation and long-range superfluid coherence. This establishes interaction-induced quasiperiodicity as a new organizing mechanism for stabilizing quasicrystalline quantum phases of matter.



Several state-of-the-art experimental platforms now offer promising routes to realize interaction-driven quasiperiodic systems. In Rydberg-atom arrays, quasiperiodic interaction profiles can be engineered by deterministic spatial patterning and laser dressing of van der Waals couplings. Trapped-ion chains provide an alternative and highly tunable setting in which spin–spin interactions mediated by collective phonon modes can be shaped with programmable optical fields (or tweezers) to realize incommensurate coupling matrices, an approach that has been explicitly proposed and is compatible with existing controls~\cite{Espinoza2021programmable,Monroe2021ions_review}. In multimode cavity-QED systems, structured global couplings naturally arise from interference between cavity modes; by designing the cavity-mode spectrum or pump geometry to enforce incommensurability, effective Fibonacci-like interaction kernels can, in principle, be implemented, with key ingredients already demonstrated~\cite{PhysRevX.8.011002,MendozaCoto2022multimodeTheory}. Furthermore, cold atoms in optical lattices coupled to structured light fields offer indirect yet flexible strategies for interaction engineering.

These experimentally established capabilities and near-term extensions indicate that the phases identified here are not merely theoretical curiosities but are within reach of current quantum-simulation platforms. Hallmark signatures of the quasi-supersolid phase, like finite superfluid density and incommensurate density modulations, can be probed with standard techniques, including time-of-flight interferometry for phase coherence and site-resolved quantum-gas microscopy (or Bragg spectroscopy) for the sharp incommensurate peaks in $S(q)$.

Looking forward, several important directions remain open. Exploring the thermal stability and melting behavior of these phases at finite temperature could reveal analogs of classical quasicrystals with quantum coherence. Investigating their real-time dynamics, response to quenches, and interaction with external drives may shed light on nonequilibrium phenomena in incommensurate settings. Finally, extending the current model to higher dimensions or multicomponent systems could uncover further connections between quasiperiodicity, supersolidity, and strongly correlated quantum matter.

\noindent\textbf{Acknowledgments--}
This work was carried out primarily at the Department pf Physics, UMASS Amherst, whose support and hospitality are gratefully acknowledged.


\bibliography{quasi}

\clearpage
\twocolumngrid   
\onecolumngrid   
\begin{center}
\textbf{\large Supplemental Material for\\
“Interaction-Induced Quasicrystalline Order: Emergence of Quasi-Solid and Quasi-Supersolid
Phases"}
\end{center}
\vspace{1em}

\setcounter{section}{0}
\renewcommand{\thesection}{S\arabic{section}}

\section{Method and Observables}

\noindent
We perform large-scale quantum Monte Carlo (QMC) simulations using the worm algorithm, which is particularly efficient for computing thermodynamic properties of strongly correlated bosonic systems. All simulations are carried out in one dimension with periodic boundary conditions (PBC), on lattices of size \( L = 55, 89, 144, 233 \) up to $987$, chosen as consecutive Fibonacci numbers to match the underlying quasiperiodic modulation. We use inverse temperature \( \beta = L \) to ensure convergence to the ground-state regime. The observables we measured are:

\vspace{4pt}
\noindent
\textit{Filling and Compressibility.}  
The average filling (density) is computed as
\begin{equation}
\langle n \rangle = \frac{1}{L} \sum_x \langle n_x \rangle,
\end{equation}
while the compressibility is given by the particle-number fluctuation estimator:
\begin{equation}
\kappa = \frac{\beta}{L} \left( \langle N^2 \rangle - \langle N \rangle^2 \right),
\end{equation}
with total particle number \( N = \sum_x \langle n_x \rangle \). Vanishing \( \kappa \) signals incompressibility.
We also perform an independent Fourier analysis of the (imaginary-time--averaged) real-space density profile,
$\langle n_x \rangle \equiv \frac{1}{\beta}\int_0^\beta d\tau\, \langle n_x(\tau)\rangle$.
Specifically, we compute
$\tilde n(q)=\frac{1}{L}\sum_x e^{iq x}\big(\langle n_x \rangle -\langle n\rangle\big)$.

\vspace{4pt}
\noindent
\textit{Superfluid Density.}  
The superfluid density \( \rho_s \) is calculated via winding numbers:
\begin{equation}
\rho_s = \frac{ \langle W^2 \rangle }{2 \beta t},
\end{equation}
where \( W \) is the spatial winding number. Finite \( \rho_s \) indicates global phase coherence, while \( \rho_s = 0 \) corresponds to insulating or localized behavior.

\vspace{4pt}
\noindent
\textit{Density-Density Correlation Function.}  
To probe spatial density order, we compute the time-averaged connected density-density correlator:
\begin{equation}
C(x) = \frac{1}{L} \sum_i \left( \langle n_i n_{i+x} \rangle - \langle n_i \rangle \langle n_{i+x} \rangle \right) - \langle n \rangle,
\end{equation}
with average density \( \langle n \rangle = \frac{1}{L} \sum_x \langle n_x \rangle \).

\vspace{4pt}
\noindent
\textit{Static Structure Factor and Fibonacci Wavevectors.}  
The static structure factor is defined as the Fourier transform of the density–density correlation function \( C(x) \):
\begin{equation}
S(q) = \frac{1}{L} \sum_x e^{i q x} C(x).
\end{equation}
Pronounced peaks in \( S(q) \) indicate emergent density order. Due to the Fibonacci-modulated long-range interactions, the leading Fourier components occur at
\begin{equation}
q^* = 2\pi \frac{F_{k-2}}{F_k}, \quad \text{and} \quad 2\pi - q^*,
\end{equation}
for a system of size \( L = F_k \), where \( F_k \) is the \( k \)-th Fibonacci number. These wavevectors arise from the Fourier transform of the characteristic function of the Fibonacci sequence and reflect the singular-continuous diffraction spectrum of quasiperiodic structures.

In particular, for \( L = 233 = F_{13} \), the dominant peaks appear at \( q^* = 2\pi \cdot 89/233 = 2.398 \) and its mirror \( 2\pi - q^* =3.882 \), consistent with the main features observed in $S(q)$. This confirms that the emergent density order is intrinsically locked to the inverse golden ratio \( \alpha = (1 - \sqrt{5})/2 \) via its rational approximants \( F_{k-2}/F_k \).

In addition to these dominant peaks, we also observe subleading features at \( q = 1.483 \) and \( 4.800 \), which correspond to higher-order harmonics at \( 2\pi \cdot F_{k-3}/F_k = 2\pi \cdot 55/233 \) and its mirror. Together, the presence of both dominant and subleading peaks confirms that the system exhibits quasiperiodic density order governed by multiple competing incommensurate wavevectors. Moreover, the dominant peak positions extracted from $S(q)$ coincide with those obtained from the Fourier spectrum of the imaginary-time--averaged density map $|\tilde n(q)|$,
confirming that these peaks reflect genuine density modulations rather than artifacts of a
particular estimator.

\vspace{4pt}
\noindent
\textit{Density Maps.}  
We also analyze the time-averaged site-resolved density \( \langle n_x \rangle \) to visualize real-space modulation patterns. In density map, each circle radii is proportional to the local density at each site.

\begin{table*}[t]
\centering
\caption{Comparison of typical observables in various quantum phases. Quasi-solid (qS) and quasi-supersolid (qSS) phases arise from incommensurate, quasiperiodic interaction modulation.}
\begin{tabular}{lccccc}
\hline \hline
Phase & $\langle n_x \rangle$ & $\kappa$ & $\rho_s$ & $C(x)$ & $S(q)$ \\
\hline
SF   & Uniform & Finite & Finite & Power-law decay & No sharp peak \\
MI   & Uniform, integer & Zero & Zero & Exponential decay & Featureless \\
CDW  & Periodic & Zero & Zero & Oscillatory with constant amplitude
 & Bragg peak at $\pi$ \\
SS   & Periodic & Finite & Finite & Oscillatory with constant amplitude
 & Bragg peak at $\pi$\\
qS   & Quasiperiodic & Zero & Zero & Incommensurate Oscillations& Sharp peaks at incomm. $q^*$ \\
qSS  & Quasiperiodic & Finite & Finite & Incommensurate Oscillations& Peaks at Incomm. $q^*$ \\
\hline \hline
\end{tabular}
\label{tab:phases}
\end{table*}
Within this system, we identify three distinct phases: the superfluid (SF), quasi-solid (qS), and quasi-supersolid (qSS) phases, each stabilized in different regions of parameter space. These phases exhibit markedly different physical properties, arising from the interplay between kinetic energy, chemical potential, and the quasiperiodic long-range interaction.

To classify and distinguish the various phases in our model, we analyze a set of key observables that encode both density and coherence properties: the local density profile $\langle n_x \rangle$, the compressibility $\kappa$, the superfluid density $\rho_s$, the time-averaged connected density-density correlation function $C(x)$, and the static structure factor $S(q)$. These quantities, together with their qualitative signatures across different phases, are summarized in Table~\ref{tab:phases}. 

In the superfluid (SF) phase, bosons delocalize uniformly across the lattice, resulting in a flat density profile, finite values of $\kappa$ and $\rho_s$, and a power-law decay of $C(x)$ reflecting long-range phase coherence. Correspondingly, $S(q)$ lacks any sharp features, consistent with the absence of translational symmetry breaking. The Mott insulator (MI) phase emerges at integer fillings under strong on-site repulsion. It is characterized by a uniform and quantized $\langle n_x \rangle$, vanishing compressibility and superfluid density, and exponential decay in $C(x)$ due to a finite gap. The structure factor remains featureless, reflecting the lack of density ordering. In contrast, the charge-density-wave (CDW) phase exhibits periodic modulations in the density profile—typically at commensurate wavevectors such as $q = \pi$—induced by nearest-neighbor or longer-range repulsions. Here, $C(x)$ shows undamped oscillations with a fixed period, $\kappa$ and $\rho_s$ both vanish, and $S(q)$ features Bragg peaks at commensurate momenta, indicating long-range crystalline order. When phase coherence coexists with such periodic order, a supersolid (SS) state arises: both $\kappa$ and $\rho_s$ are finite, $C(x)$ maintains long-range oscillatory behavior, and $S(q)$ simultaneously hosts a sharp Bragg peaks at $q = \pi$ or other rational wavevectors.

Distinct from these commensurate states, the quasi-solid (qS) phase in our system is stabilized purely by the deterministic incommensurability of the long-range interactions. The spatial density profile $\langle n_x \rangle$ develops quasiperiodic modulations that follow a non-repeating yet deterministic pattern, tied to the irrational modulation parameter $\alpha = (\sqrt{5}-1)/2$. These modulations break translational symmetry without introducing disorder. In the qS phase, both $\kappa$ and $\rho_s$ vanish, indicating an incompressible, non-superfluid state, while $C(x)$ exhibits oscillatory correlations. This is most clearly captured in the structure factor $S(q)$, which reveals a series of sharp peaks located at irrational wavevectors such as $q^* = 2\pi \cdot 89/233$, $2\pi \cdot 55/233$, and so on. These wavevectors correspond to dominant Fourier components of the density profile and are directly determined by the Fibonacci-derived modulation of the interaction matrix. Their irrational character rules out conventional commensurate ordering and confirms the quasiperiodic origin of the spatial structure. Finally, the quasi-supersolid (qSS) phase retains the aperiodic density modulations of the qS state but also supports finite superfluid superfluid $\rho_s > 0$, indicating emergent phase coherence. In this phase, $C(x)$ displays incommensurate oscillations, and $S(q)$ exhibits a set of sharp incommensurate peaks. This coexistence of superfluid response and deterministic incommensurate density order, in the absence of external potentials or commensurate lattices, demonstrates the genuinely new character of the qSS phase—fundamentally distinct from supersolids with periodic order.

\begin{figure*}[t]
    \centering
    \includegraphics[width=0.6\linewidth]{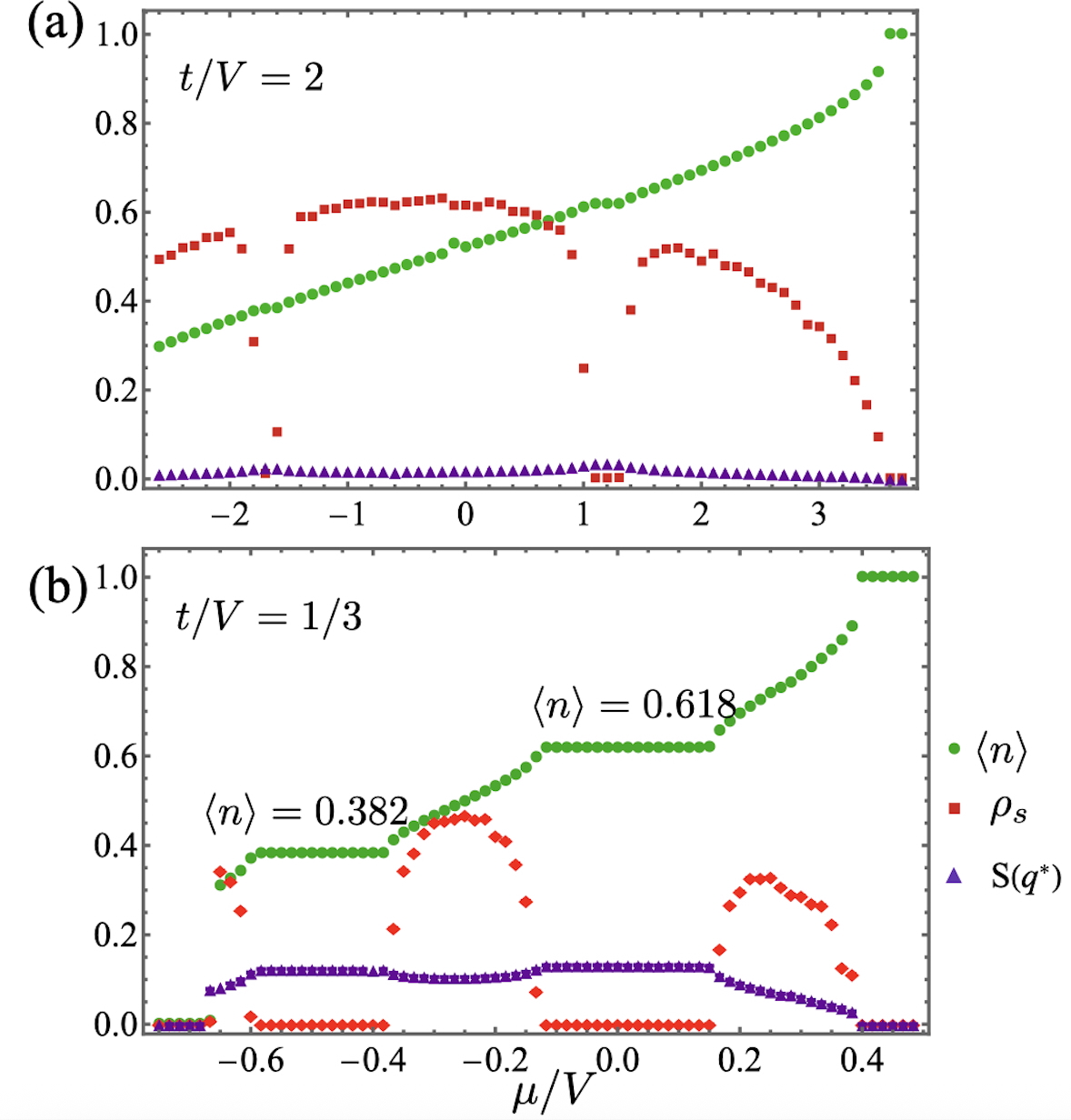}
    \caption{
    (a) Average filling $\langle n \rangle$, superfluid density $\rho_s$, and structure factor $S(q^*)$ 
    as functions of the chemical potential $\mu/V$ for $t/V = 2$ at system size $L = 233$. 
    The system is predominantly superfluid, exhibiting finite $\rho_s$, 
    while $S(q^*)$ remains nearly zero across most of the chemical potential range. 
    A weak enhancement of $S(q^*)$ ($\max S(q^*) \approx 0.035$) occurs near the inverse golden-ratio fillings 
    $\langle n \rangle = 0.382$ and $\langle n \rangle = 0.618$, indicating the quasiperiodic order.  
    (b) Corresponding data for $t/V = 1/3$.  
    In this regime, well-defined incompressible plateaus appear at 
    $\langle n \rangle= 0.382$ and $\langle n \rangle = 0.618$, accompanied by enhanced $S(q^*) $, 
    marking the emergence of stable quasi-solid phases.  
    Between the plateaus, $\rho_s$ remains finite while $S(q^*)$ stays nonzero, 
    signifying a quasi-supersolid region where phase coherence coexists with incommensurate density modulation.  
    }
    \label{fig:rho_mu}
\end{figure*}

\section{Additional $\mu/V$–$n$ Curves at Different $t/V$}

To further verify the robustness of the quasi-solid and quasi-supersolid phases reported in the main text, we present additional results for the particle density $\langle n \rangle$, the superfluid density $\rho_s$, and the static structure factor $S(q^*)$, plotted as functions of the chemical potential $\mu/V$ at two representative interaction strengths: $t/V= 2$ and $t/V = 1/3$. All simulations are performed using the worm algorithm at inverse temperature $\beta = L$ for a system size $L = 233$.

\smallskip
\noindent\textbf{Identification of Quasi-Periodic Order--}
To quantitatively distinguish quasi-periodic density modulations, we introduce a practical criterion based on the structure factor peak value at the dominant incommensurate wavevector $q^*$. This wavevector corresponds to the main Fourier component of the quasiperiodic interaction pattern and determines the modulation wave number of the emergent quasi-solid density order. For the Fibonacci-modulated system considered here, the characteristic momentum satisfies
\begin{equation}
    q^* = 2\pi \frac{F_{k-2}}{F_k},
\end{equation}
where $F_k$ denotes the $k$-th Fibonacci number, chosen to match the system size $L = F_k$, and $F_{k-2}$ is the preceding Fibonacci number. For the present lattice size $L = 233 = F_{13}$ and $F_{11} = 89$, we thus obtain $q^* = 2\pi \times 89/233 \simeq 2.398$, in excellent agreement with the dominant peaks observed in $S(q)$ (see main text, Fig.~3).


To identify the presence of quasiperiodic density order, we adopt a conservative threshold criterion based on the peak height of the structure factor. Specifically, quasi-solid correlations are considered significant when $S(q^*) > 0.03$, while the system is regarded as featureless for $S(q^*) < 0.01$. The intermediate regime, $0.01 < S(q^*) < 0.03$, corresponds to a region where incommensurate correlations remain but are too weak to support long-range order. This criterion is empirically motivated: as the quasi-solid plateau vanishes, $S(q^*)$ drops to around 0.01. Statistically, the threshold is robust, since the typical Monte Carlo uncertainty in $S(q^*)$ is below $2 \times 10^{-4}$—more than an order of magnitude smaller than the weakest quasi-solid signal we identify.

\smallskip
\noindent\textbf{Weak Interaction Regime ($t/V = 2$)--}
Figure~S1(a) shows the behavior of $\langle n \rangle$, $\rho_s$, and $S(q^*)$ as functions of $\mu/V$ in the weak-interaction regime. The system is predominantly superfluid: $\rho_s$ is finite, and $S(q^*)$ stays very small over most of the $\mu/V$ range. However, near fillings $\langle n \rangle =0.618 \approx \alpha$ and $0.382=1-\alpha$, $S(q^*)$ develops a shallow maximum (peaking at $\sim 0.035$) and $\langle n \rangle(\mu/V)$ exhibits narrow plateaus. According to our conservative criterion ($S(q^*)>0.03$), these windows mark the onset of quasi-solid correlations and incipient incompressibility, identifying them as quasi solid regions. Outside these plateaus, $S(q^*)$ quickly falls below 0.03, indicating that the system remains a compressible superfluid.

\smallskip
\noindent\textbf{Strong Interaction Regime ($t/V= 1/3$)--}
Figure~S1(b) presents the results for stronger interaction. In this regime, clear incompressible plateaus emerge at fillings $\langle n \rangle = 0.382$ and $\langle n \rangle = 0.618$, both accompanied by sharply enhanced values of $S(q^*)$ exceeding the quasi-solid threshold. These plateaus correspond to stable quasi-solid phases whose incommensurate density modulations are locked to the golden-ratio wavevector $q^*$. Between the plateaus, $\rho_s$ remains finite while $S(q^*)$ stays nonzero ($S(q^*) > 0.03$), indicating the coexistence of phase coherence and quasiperiodic density modulation—signatures of a quasi-supersolid state.

\section{Finite-Size Scaling of Structure Factor and Superfluid Density}
\label{sec:scaling}

\begin{figure}[t]
    \centering
    \includegraphics[width=0.48\textwidth]{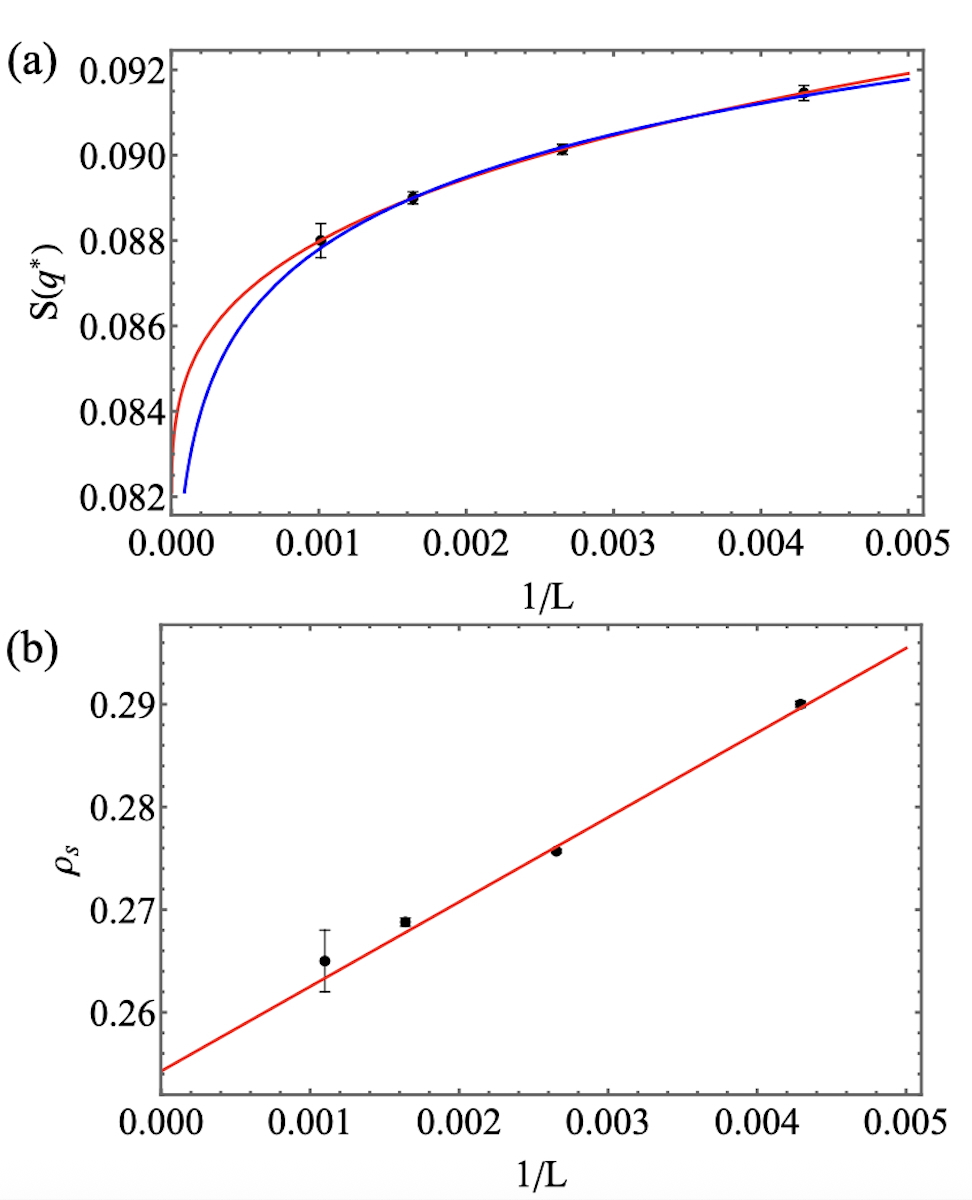}
    \caption{Finite-size scaling of (a) the structure factor peak $S(q^*)$ and (b) the superfluid density $\rho_s$ in the quasi-supersolid regime at filling $\langle n \rangle=0.694$. 
    In (a), both constant-like (red) and power-law (blue) fits are shown; the saturation of $S(q^*)$ indicates quasi-long-range density order. 
    In (b), the linear fit of $\rho_s$ versus $1/L$ extrapolates to a finite value, demonstrating robust phase coherence in the thermodynamic limit.}
    \label{fig:sq_scaling}
\end{figure}

To confirm the stability of the quasi-supersolid phase in the thermodynamic limit, we perform finite-size scaling analyses of both the structure factor at the incommensurate wavevector, $S(q^*)$, and the superfluid density $\rho_s$. 

Figure~\ref{fig:sq_scaling}(a) shows the finite-size scaling of the dominant incommensurate peak $S(q^*)$ of the static structure factor at filling $\langle n \rangle = 0.694$. Two types of fitting functions are employed to probe the thermodynamic behavior. The red curve represents a ``constant-like'' form, $S(q^*)(L) = a + b\,L^{-c}$, with best-fit parameters $a = 0.0814$, $b = 0.0494$, and $c = 0.2928$, implying convergence to a finite value in the large-$L$ limit. The blue curve shows a power-law fit, $S(q^*)(L) = d\,L^{-e}$, with $d = 0.1062$ and a very small exponent $e = 0.0275$, indicating an extremely slow decay.

Both fits reproduce the data well within Monte Carlo uncertainties. The near-flatness of the power-law decay and the quantitative agreement with the constant-like form strongly suggest that $S(q^*)$ saturates to a nonzero value as $L \to \infty$. This is consistent with theoretical expectations of finite Bragg-like peaks in Fibonacci-modulated systems, and supports the existence of robust quasiperiodic density modulations. Taken together, the results indicate that the quasi-supersolid phase remains stable in the thermodynamic limit.

In Fig.~\ref{fig:sq_scaling}(b) we show the finite-size scaling of the superfluid density $\rho_s(L) = a+b/L$ with $a=0.254278$ and $b=8.23709$, extracted from winding-number fluctuations. The data are well described by a linear extrapolation in $1/L$, which approaches a nonzero constant in the $L\to\infty$ limit. This confirms the persistence of global phase coherence.

Taken together, these results establish that the quasi-supersolid phase is not a finite-size artifact but survives in the thermodynamic limit, characterized by the coexistence of incommensurate density modulations ($S(q^*)>0$) and finite superfluid density.

\section{Correlation Functions at Large System Size}
\label{sec:largeL}

\begin{figure}[t]
    \centering
    \includegraphics[width=0.7\linewidth]{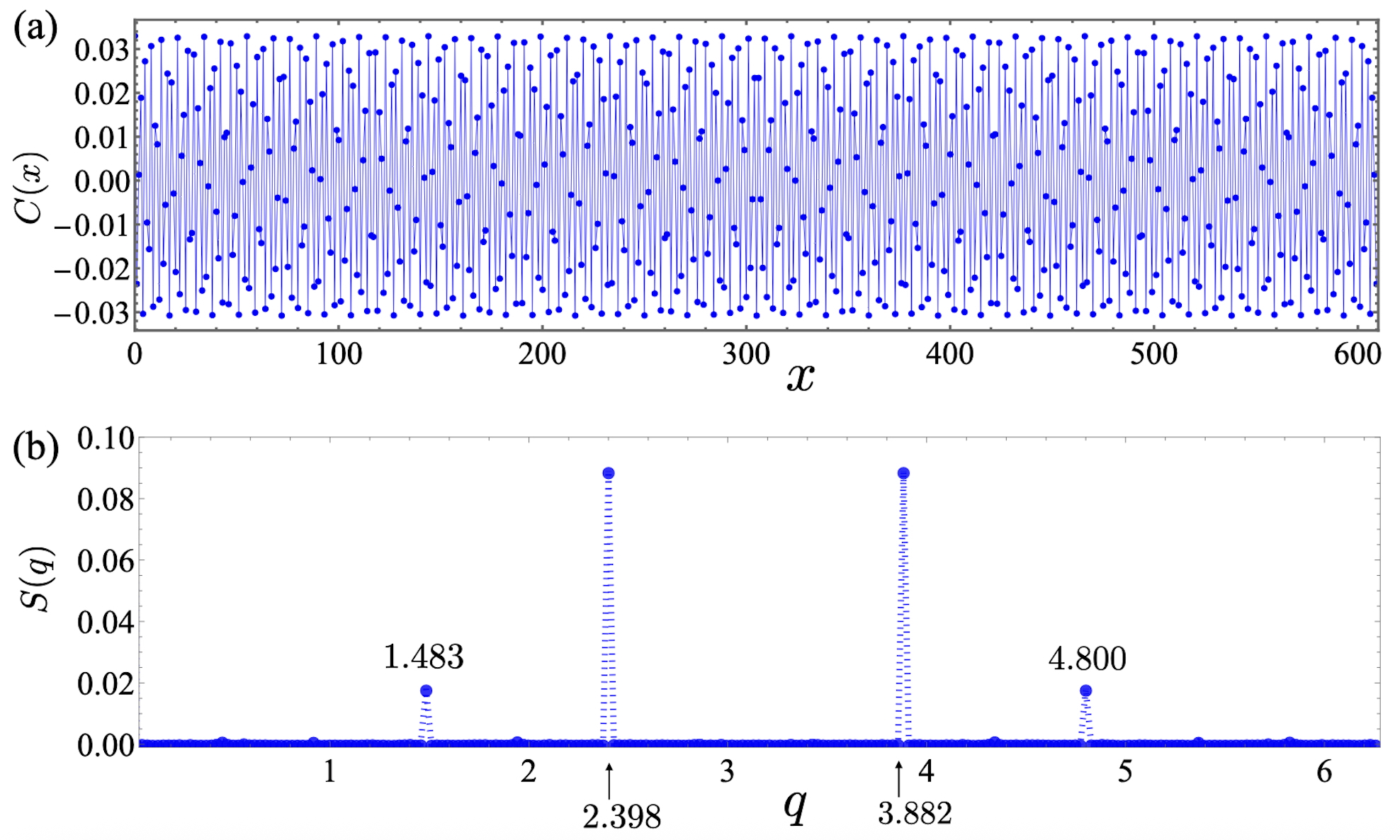}
    \caption{
    (a) Real-space density–density correlation function $C(x)$ for a large system of size $L=610$ at interaction strength $t/V=1/3$ and filling $\langle n \rangle = 0.694$. The correlation exhibits long-range quasiperiodic oscillations across the entire system, confirming the persistence of incommensurate order. 
    (b) Corresponding structure factor $S(q)$ showing sharp peaks at incommensurate wavevectors $q^* = 1.483$, $2.398$, $3.882$, and $4.800$, directly associated with Fibonacci ratios.    }
    \label{fig:large_system}
\end{figure}

To further verify the robustness of the quasi-solid and quasi-supersolid phases identified in the main text, we perform additional simulations on a substantially larger system with $L=610$. Figure~\ref{fig:large_system} presents the equal-time density–density correlation function $C(x)$ and its Fourier transform $S(q)$ at a filling $\langle n \rangle = 0.694$, under the interaction strength $t/V = 1/3$.

The equal-time connedted density-density correlation $C(x)$ exhibits clear, long-range oscillations spanning the entire chain, indicating that quasiperiodic density order remains coherent even at this system size. To quantify this order, we fit $C(x)$ using the physically motivated form:
\begin{equation}
    C(x) = A \cos(q x) e^{-x/\xi} + C_0,
\end{equation}
which models a modulated density pattern with dominant wavevector $q$ and an exponential decay governed by correlation length $\xi$. 

For this dataset, we obtain the following fit parameters:
\begin{equation*}
    q = 2.39997, \quad A = 0.0361, \quad \xi = 1.085 \times 10^6, \quad C_0 = 2.20 \times 10^{-8}.
\end{equation*}
The fitted $q$ value matches the dominant incommensurate peak observed in $S(q)$ near $q^* = 2.398$, which is consistent with a Fibonacci-induced modulation. The extremely large correlation length $\xi$ indicates that the spatial order is maintained over essentially the full system length, demonstrating the thermodynamic stability of quasi-solid correlations.

The structure factor $S(q)$ in Fig.~\ref{fig:large_system}(b) reinforces this picture, showing sharp Bragg-like peaks at incommensurate positions such as $q^* = 1.483$, $2.398$, $3.882$, and $4.800$, all of which are linked to characteristic Fibonacci ratios. These features reflect the presence of multiple competing periodicities in the system’s density modulation spectrum, stabilized by the underlying quasiperiodic interaction.

Specifically, the dominant peak at $q^* = 2.398$ corresponds to the primary Fourier component of the Fibonacci modulation at $q^* = 2\pi F_{k-2}/F_k = 2\pi \cdot 89 / 233$, where $F_k$ denotes the $k$-th Fibonacci number and $L = 233 = F_{13}$ in our case. Its mirror-symmetric counterpart appears at $q = 3.882 \approx 2\pi - q^*$, consistent with the intrinsic inversion symmetry of the modulation. In addition, two weaker peaks emerge at $q^* = 1.483$ and $4.800$, which correspond to higher-order Fourier components. The former arises from the ratio $2\pi \cdot F_{k-3}/F_k = 2\pi \cdot 55 / 233$, and the latter is its mirror image $2\pi - q^*$. These subleading peaks reflect the self-similar, hierarchical structure of the Fibonacci modulation and indicate the presence of multiple rational approximants of the inverse golden ratio $\alpha = (1 - \sqrt{5})/2$ encoded in the density response. Together, the appearance of these four peaks confirms the quasiperiodic nature of the emergent order and matches expectations from the mathematical theory of aperiodic diffraction.

Importantly, the same parameter point also exhibits finite superfluid response. The measured superfluid density is $\rho_s = 0.2688 \pm 0.0005$, indicating that the system supports both coherent density modulations and phase coherence. This combination of long-range incommensurate density order and superfluidity places the system unambiguously in the quasi-supersolid phase.

Together with the finite-size scaling results discussed before, these large-system data confirm that the quasi-supersolid phase is not a finite-size artifact, but a genuine many-body state stabilized by the interplay of long-range interactions and quantum fluctuation. While the filling $\langle n \rangle= 0.694$ is slightly off the Fibonacci ratio $\langle n \rangle= 0.618 \approx \alpha$, the persistence of both superfluidity and long-range order underscores the robustness of quasi-supersolid phase across a broad parameter regime.

\end{document}